\documentclass[twocolumn,showpacs,preprintnumbers,amsmath,amssymb,aps]{revtex4}
\usepackage{graphicx}
\usepackage{dcolumn}
\usepackage{bm}
\begin{document}
\preprint{}
\title{On attributes of a Rotating Neutron star with a Hyperon core}
\author{T. K. Jha{\footnote {email: tkjha@prl.res.in}},
H. Mishra{\footnote {email: hm@prl.res.in}} and 
V. Sreekanth{\footnote {email: skv@prl.res.in}}}
\affiliation { Theoretical Physics Division, Physical Research Laboratory, Navrangpura, 
Ahmedabad, India - 380 009}
\date{\today}
\begin{abstract}

We study the effect of rotation on global properties of neutron star with a
hyperon core in an effective chiral model with varying nucleon effective mass 
within a mean field approach. The resulting gross properties of the rotating compact star sequences are then compared and analyzed with other theoretical 
predictions and observations from neutron stars. The maximum mass of the compact star 
predicted by the model lies in the range $(1.4-2.4) ~M_{\odot}$ at Kepler frequency $\Omega_K$, 
which is consistent with recent observation of high mass stars thereby reflecting 
the sensitivity of the underlying nucleon effective mass in the dense matter EoS. We also
discuss the implications of the experimental constraints from the flow data from heavy-ion
collisions on the global properties of the rotating neutron stars.

\end{abstract}
\maketitle
\section{Introduction}

Recent advances in the study of cold dense matter has received new impetus
both from observational astronomy and the recent heavy-ion collision experiment as well as the upcoming compressed baryonic matter (CBM) experiments, that can put constraints 
on the equation of state (EoS) of dense matter \cite{a}. Arguably the uncertainties 
in the dense matter EoS can also be reduced appreciably through theoretical modeling of neutron stars and by analyzing their global properties with an eye on the
observational data from these compact objects. Consequently, building an EoS 
in accordance with the heavy-ion collision data and its subsequent application
to the dense matter problem seems to be prerequisite. The study
can throw significant light on potential problems such as the isospin dependence 
of nuclear forces, the presence of stable hyperon in neutron star core and its
impact on cold dense matter \cite{A1,mr}.

Lately, there has been some striking signatures or observations of both massive and 
the low mass stars, which may have interesting implications on the structure and
composition of neutron stars \cite{baym}. For example, the observation of massive compact
stars $(M \approx 2 M_{\odot})$ \cite{mass,mass1} from observation of QPO's (Quasi Periodic 
Oscillations) from X-ray emmisions on one hand and also the least massive ones 
$1.18 \pm 0.02 M_{\odot}$ \cite{least} from the binary pulsar $J1756-2251$ emphasize the
need to address key issues in limiting the observed global properties of the stars
in th $M-R$ plane. On the contrary, the canonical value of
$M = 1.44 M_{\odot}$, the largest precisely measured mass of the Neutron star $PSR 1913+16$ 
\cite{exact} effectively doesn't seem to provide any stringent condition and
most of the existing realistic equation of state satisfies the criteria.
These aforesaid facts signifies the need to constraint the nuclear 
EoS of dense matter which are extrapolations of the properties defined 
in the vicinity of nuclear saturation density and therefore it is important
to test these extrapolations with regard to the available experimental
flow data from heavy-ion collisions. 

Equivalently considerable emphasis is being laid to understand nuclear matter EoS in the 
density range of ($1-3$)$\rho_0$, the region from where the entire 
possibilities of exotic forms of matter (hyperons, condensates of bosons or quarks) 
\cite{form,form1,form2,form3,form4} starts appearing. It is 
known that the neutron star radius is sensitive to the properties of nuclear force in the 
immediate vicinity of nuclear saturation density $(\rho_0)$, whereas the 
maximum mass of the star is regulated by the dense matter EoS well beyond 
$2\rho_0$ \cite{form4}. Among the various scenarios related to the star structure, 
the presence of stable hyperons is an exciting possibility 
and is known to have significant impact on the dense matter 
equation of state and in turn gets manifested in
the star properties too \cite{nkg,prak97,form1}. The hyperons
such as $\Lambda^{0}$(1116), $\Sigma^{-,0,+}$(1193) and $\Xi^{-,0}$(1318) 
appear at the expense of nucleons and forms a sizeable population in the 
density region of current interest. Although hypernuclei experiments
\cite{hyper} supports the presence of hyperons in dense matter, 
however this domain is also marred by uncertainties revolving around the 
nature of hyperon-hyperon interactions and 
their coupling strengths. Since they are the dominant species in the dense 
matter population, there is even greater need to understand the nature and 
behavior of these interactions. 

Apart from that, it is also known that the addition of
quartic terms softens the EoS considerably in addition to that caused by the hyperons,
which thereby makes it difficult to obtain a neutron star mass 
larger than $2 M_{\odot}$ \cite{ser96}. Here in the present investigation, we 
employ a model which has chiral symmetry and also embodies a dynamical generation 
of the vector meson mass along with the non-linear term in the scalar 
field interaction. The model has been used extensively to dense matter studies 
\cite{prakash,five,dutta}, which was then modified with the addition of non-linear terms in 
the scalar field interaction to reproduce nuclear saturation properties 
at reasonable incompressibility \cite{sahu,tkj04}. Finally we generalized the model
to include the lowest lying octet of baryons in our previous work \cite{tkj06} and applied
it to study neutron star matter with varying incompressibility and varying 
nucleon effective mass. From our previous analysis \cite{tkj06}, it was apparent that
the difference in the nuclear incompressibility was neither prominent in the
EoS, nor in the static star properties, whereas the EoS was extremely sensitive to the 
underlying nucleon effective masses and hence was also visible in the global properties
of the star. In continuation of our earlier work, we now extend our analysis 
to study the rotational attributes of neutron star with varying nucleon 
effective mass in the mean-field approach within the same hadronic model. 
In short, our motivation here has been to investigate how the rotational 
attributes of the neutron star restrict the parameters of the present model. 
 
The outline of the paper is as follows: In the next section, we give a brief description
of the ingredients of the hadronic model that we implement in our present 
work. After describing the numerical scheme for calculating the 
rotating neutron star sequences, the general features of the 
equation of state of hyperon rich matter is studied and the global
properties of the rotating neutron star is presented. We shall 
analyze these results with respect to varying nucleon effective mass
and also talk of a few constraints on the neutron star mass and radius
imposed by recent observations. Finally we conclude with outlook on the 
possible extensions of the current approach.

\section{The equation of state}
In this context, we tried to explore the consequences of an EoS based on chiral
$\sigma$ model with nonlinear interaction terms in the scalar field. 
To have a realistic description of the dense neutron 
star matter, we consider the effective Lagrangian of the chiral model generalized 
to include the lowest lying octet of baryons ($n,p,\Lambda^{0},\Sigma^{-,0,+},\Xi^{-,0}$) 
interacting through the exchange of the pseudo-scalar meson $\pi$, the scalar 
meson $\sigma$, the vector meson $\omega$ and the iso-vector $\rho-$meson, 
and is given by \cite{tkj06}:

\begin{widetext}
\begin{eqnarray}
\label{lag}
{\cal L}&=& \bar\psi_B~\left[ \big(i\gamma_\mu\partial^\mu
         - g_{\omega B}\gamma_\mu\omega^\mu
         - \frac{1}{2}g_{\rho B}{\vec \rho}_\mu\cdot{\vec \tau}
            \gamma^\mu\big )
         - g_{\sigma B~}~\big(\sigma + i\gamma_5
             \vec \tau\cdot\vec \pi \big)\right]~ \psi_B
\nonumber \\
&&
        + \frac{1}{2}\big(\partial_\mu\vec \pi\cdot\partial^\mu\vec\pi
        + \partial_{\mu} \sigma \partial^{\mu} \sigma\big)
        - \frac{\lambda}{4}\big(x^2 - x^2_0\big)^2
        - \frac{\lambda B}{6}\big(x^2 - x^2_0\big)^3
        - \frac{\lambda C}{8}\big(x^2 - x^2_0\big)^4
\nonumber \\
&&      - \frac{1}{4} F_{\mu\nu} F_{\mu\nu}
        + \frac{1}{2}{g_{\omega B}}^{2}x^2 \omega_{\mu}\omega^{\mu}
        - \frac {1}{4}{\vec R}_{\mu\nu}\cdot{\vec R}^{\mu\nu}
        + \frac{1}{2}m^2_{\rho}{\vec \rho}_{\mu}\cdot{\vec \rho}^{\mu}\ .
\end{eqnarray}
\end{widetext}

The first line of the above Lagrangian represents the interaction of baryons $\Psi_B$
with the aforesaid mesons. In the second line we have the kinetic and the
non-linear terms in the pseudoscalar-isovector pion field `$\vec \pi$',  
the scalar field `$\sigma$', and with $x^2= {\vec \pi}^2+\sigma^{2}$.
Finally in the last line, we have the field strength and the mass term for the vector
field `$\omega$' and the iso-vector field `$\vec \rho$' meson.
The terms in eqn. (1) with the subscript $'B'$ 
should be interpreted as sum over the states of all baryonic 
octets. In this paper we shall be concerned only with the normal non-pion condensed
state of matter, so we take $<\vec \pi>=0$.

The interaction of the scalar and the pseudoscalar mesons with the vector 
boson generates a dynamical mass for the vector bosons through 
spontaneous breaking of the chiral symmetry with scalar field getting the 
vacuum expectation value $x_0$. Then the masses of the baryons, the scalar and the 
vector mesons, are respectively given by

\begin{eqnarray}
m_B = g_{\sigma B} x_0,~~ m_{\sigma} = \sqrt{2\lambda} x_0,~~
m_{\omega} = g_{\omega B} x_0\ .
\end{eqnarray}

In the above, $x_0$ is the vacuum expectation value of the $\sigma$ field.
We could have taken an interaction of the $\rho-$meson with the scalar
and the pseudoscalar mesons similar to the omega meson. However, a dynamical mass
generation mechanism of the $\rho-$meson in a similar manner will not generate 
the correct symmetry energy. Therefore, we have taken an explicit mass term
for the isovector $\rho-$meson similar to what was considered in earlier 
works \cite{dutta,sahu,tkj04,tkj06}.
 
We employ the mean-field procedure to evaluate the meson fields in our present
calculations. In the mean-field treatment, one assumes the mesonic fields to be uniform 
i.e., without any quantum fluctuations. We recall here that this approach has been 
extensively used to obtain field-theoretical EoS for high density matter \cite{five}, 
and gets increasingly valid when the source terms are large \cite{serot86}.
The details of the model that we use in our present investigation and its
attributes such as the derivation of the equation of motion of 
the meson fields and its equation of state $(\varepsilon~\&~P)$
of the many baryonic system, can be found in our preceding work \cite{tkj06}.
For the sake of completeness however we write down the meson field equations. 
The vector and the iso-vector fields are respectively given by

\begin{eqnarray}
\omega_0=\sum_{B}\frac{ \rho_B }{g_{\omega B} x^2} \ ,
\end{eqnarray}

\begin{eqnarray}
\rho_{03} =\sum_{B} \frac{g_{\rho B}}{m_\rho^2} I_{3 B}\rho_{B}\ .
\end{eqnarray}

In the above equations the quantity $\rho_B$ is the Baryon density 
and $I_{3 B}$ is the 3rd-component of the isospin of each baryon species.

The scalar field equation can be written in terms of the variable
$Y=x/x_0$ with $x=(<\sigma^2+\pi^2>)^{1/2}$ as \cite{tkj06}

\begin{widetext}
\begin{equation}
\sum_B\left[ (1-Y^2) -\frac{B}{c_{\omega B}}(1-Y^2)^2
+\frac{C}{c_{\omega B}^2}(1-Y^2)^3
+\frac{2 c_{\sigma B}c_{\omega B}\rho_B^2}{m_{B}^2Y^4}
-\frac{2 c_{\sigma B}\rho_{SB}}{m_{B} Y}\right]=0\ ,
\label{effmass}
\end{equation}
\end{widetext}
\noindent
where the effective mass of the baryonic species is $m_{B}^{\star} 
\equiv Ym_{B}$ and $c_{\sigma B}\equiv  g_{\sigma B}^2/m_{\sigma}^2 $ are the
$c_{\omega B} \equiv g_{\omega B}^2/m_{\omega}^2 $ are the usual
scalar and vector coupling constants respectively. 
It may be noted that the parameter `$\lambda$' in the
Lagrangian does not appear explicitly in eqn. (5), but enters implicitly
through the mass term for the scalar meson, following equation (2).
Similarly, in the present model describing dense matter, the $\omega-$meson 
mass is generated dynamically. This vector meson mass enters in Eq. 
(\ref{effmass}) through the ratio
$c_{\omega} = (g_{\omega}/m_{\omega})^2 \equiv 1/{x_0}^2$ which has to be fixed 
alongwith $c_{\sigma}$ so as to satisfy the nuclear matter saturation 
properties, similar to what has
been done in earlier works \cite{dutta,sahu,tkj04,tkj06}. 
Further in Eq.(\ref{effmass}), the quantities $\rho_B$ and 
$\rho_{SB}$ are the baryon density and the scalar density for a given
baryon species given respectively as,

\begin{equation}
\rho_B= \frac{\gamma}{(2\pi)^3}\int^{k_B}_o d^3k,
\end{equation}

\begin{equation}
\rho_{SB}= \frac{\gamma}{(2\pi)^3}\int^{k_B}_o\frac{m^*_{B} d^3k}
         {\sqrt{k^2+m_{B}^{\star 2}}},
\end{equation}
\noindent
where $k_B$ is the fermi momentum of the baryon and $\gamma=2$ is the spin
degeneracy factor.

We now go directly to the total energy density `$\varepsilon$' and pressure `$P$' 
for a given baryon density in terms of the dimensionless variable $Y=x/x_0$ 
which is given as:

\begin{widetext}
\begin{eqnarray}
\label{ep0}
\varepsilon
&=&
 \frac{2}{\pi^2}\int^{k_B}_0 k^{2}dk{\sqrt{k^2+m_B^{\star 2}}}
         +  \frac{m_B^2(1-Y^2)^2}{8c_{\sigma B}}
        - \frac{m_B^2 B}{12c_{\omega B}c_{\sigma B}}(1-Y^2)^3
\nonumber \\
        &+& \frac{m_B^2 C}{16c_{\omega B}^2c_{\sigma B}}(1-Y^2)^4
        + \frac{1}{2Y^2}{c_{\omega_{B}} \rho_B^2}
        +\frac{1}{2}m_{\rho}^{2}\rho_{03}^2
        + \frac{1}{\pi^{2}}\sum_{\lambda=e,\mu^{-}}\int^
           {k_\lambda}_0 k^{2}dk{\sqrt{k^2+m^2_{\lambda}}}\ ,
\end{eqnarray}
\begin{eqnarray}
P &=&
          \frac{2}{3\pi^2}\int^{k_B}_0\frac{k^{4}dk}
          {{\sqrt{k^2+m_B^{\star 2}}}}
        - \frac{m_B^2(1-Y^2)^2}{8c_{\sigma B}}
        + \frac{m_B^2 B}{12c_{\omega B}c_{\sigma B}}(1-Y^2)^3
\nonumber \\
&-&
         \frac{m_B^2 C}{16c_{\omega B}^2c_{\sigma B}}(1-Y^2)^4
        + \frac{1}{2Y^2}{c_{\omega_{B}} \rho_B^2}
        + \frac{1}{2}m_{\rho}^{2}\rho_{03}^2\
       +\frac{1}{3\pi^2}\sum_{\lambda=e,\mu^{-}}\int^{k_\lambda}_0
          \frac{k^{4}dk}{{\sqrt{k^2+m^2_{\lambda}}}}
\end{eqnarray}
\end{widetext}

The terms in eqns. (3) and (4) with the subscript $`B'$ should be interpreted 
as sum over all the states of the baryonic octets.
The meson field equations for the $\sigma$, $\omega$ and $\rho-$mesons are then solved
self-consistently at a fixed baryon density to obtain the respective field strengths.
The EoS for the $\beta-$equilibrated for the hyperon rich matter is obtained
with the requirements of conservation of total baryon number and charge
neutrality condition given by \cite{tkj06}

\begin{equation}
\sum_{B}Q_{B}\rho_{B}+\sum_{l}Q_{l}\rho_{l}=0,
\end{equation} 
\noindent
where $\rho_{B}$ and $\rho_{l}$ are the baryon and the lepton (e,$\mu$) number
densities with $Q_{B}$ and $Q_{l}$ as their respective electric charges.

Using the computed EoS for the neutron star sequences, we calculate the 
structural properties of neutron stars with a hyperons core.

\section{Stellar Equations}

The equations for the structure of a relativistic spherical and
static star composed of a perfect fluid were derived from Einstein's
equations by Tolman, Oppenheimer and Volkoff \cite{tov}, which are

\begin{equation}
\frac{dP}{dr}=-\frac{G}{r}\frac{\left[\varepsilon+P\right ]
\left[M+4\pi r^3 P\right ]}{(r-2 GM)},
\label{tov1}
\end{equation}
\begin{equation}
\frac{dM}{dr}= 4\pi r^2 \varepsilon,
\label{tov2}
\end{equation}
\noindent

with $G$ as the gravitational constant and $M(r)$ as the enclosed
gravitational mass. We have used $c=1$.
Given an EOS, these equations can be integrated from the origin as an initial
value problem for a given choice of central energy density, $(\varepsilon_c)$.
The value of $r~(=R)$, where the pressure vanishes defines the
surface of the star. We solve the above equations to study the structural properties of a
static neutron star using the EoS derived for the electrically charge
neutral hyperonic dense matter \cite{nr,nr1}.

Although a relativistic compact star has much complicated internal structure
but its properties can be reasonably approximated by some simplifying assumptions.
The matter inside is assumed to be a perfect fluid on the basis of the 
observation from pulsar glitches, which shows that the departure from 
perfect fluid equilibrium due to the solid crust is quite negligible
($\sim 10^{-5}$) \cite{fluid}. At birth, a neutron star is differentially
rotating, but because of several factors such as the cooling phenomenon,
shear viscosity, neutrino diffusion etc., the star assumes a uniform rotation.
So the approximation of a zero temperature perfect fluid neutron star matter 
is a good one. 

In order to calculate the models of rotating star basically two approaches are being
employed \cite{fried1}, namely the Hartle approach (slow rotation) \cite{hartle} and the 
Komatsu, Eriguchi, and Hachisu (KEH) method (fast rotation) \cite{kmt,kmt1} approach.
Although an improved version of the former approach was employed to 
calculate the properties of rotating
stars \cite{slow,slow1,mishra}, large descripancies has been noted compared to models
without the assumption of slow rotation, particularly near the mass-shedding limit \cite{des}. 
In our present investigation, we employ the KEH approach to calculate the model for rapidly
rotating stars near the mass-shedding limit or conversely upto the Kepler frequency, 
which we briefly describe now. 

Let us consider equilibrium stars in uniform 
rotation with static, axial symmetric space-time. Now the time translational 
invariant and axial-rotational invariant metric in spherical polar coordinates 
(\textit{t,r}, $\theta,\phi$) can be written as

\begin{equation} 
ds^2=-e^{2\nu} dt^2+e^{2\alpha} (dr^2+r^2 d\theta^2) 
+e^{2\beta}r^2\sin^2\theta (d\phi - \omega dt)^2, 
\label{metric}
\end{equation}
\noindent
where the metric functions $\nu, \alpha, \beta, \omega$ depends only on $\textit{r}$ and $\theta$. By treating stellar matter as perfect fluid, the energy momentum tensor can be written as

\begin{equation}
T^{\mu\nu} = Pg^{\mu\nu} + (P+\epsilon) u^{\mu} u^{\nu},
\end{equation}
\noindent
with the four-velocity 

\begin{equation}
u^{\mu} = \frac{e^{-\nu}}{\sqrt{1-v^2}}(1,0,0,\Omega).
\end{equation}
\noindent
Here 

\begin{equation}
\mathit{v}=(\Omega-\omega) r\sin\theta e^{\beta-\nu}, \label{vsin}
\end{equation}
\noindent
is the proper velocity relative to an observer with zero angular velocity and $\Omega$ is the angular velocity of the star measured from infinity. Now we can compute the Einstein field equations given by

\begin{eqnarray}
\mathit{R}_{\mu \nu} - \frac{1}{2} g_{\mu \nu}\mathit{R} = {8\pi}T_{\mu \nu}
\end{eqnarray}
\noindent
(where $\mathit{R}_{\mu \nu}$ is the Ricci tensor and and $\mathit{R}$ is the 
scalar curvature and with c = G = 1). This leads to the equations for the
metric functions as

\begin{eqnarray}
\Delta\left[ \rho e^\frac{\gamma}{2}\right]  = S_{\rho}(r,\mu),\\
\left( \Delta+\frac{1}{r}\frac{\partial}{\partial r}-\frac{1}{r^2}\mu \frac{\partial}{\partial \mu}\right) \gamma e^\frac{\gamma}{2} = S_{\gamma}(r,\mu),\\
\left( \Delta+\frac{2}{r}\frac{\partial}{\partial r}-\frac{2}{r^2}\mu \frac{\partial}{\partial \mu}\right) \omega e^\frac{\gamma-2\rho}{2} = S_{\omega}(r,\mu),
\end{eqnarray}
\noindent
where

\begin{eqnarray*}
\Delta = \frac{\partial^2}{\partial r^2}+\frac{2}{r}\frac{\partial}{\partial r}+\frac{1}{r^2}\frac{\partial^2}{\partial \theta^2}+\frac{1}{r^2}\cot\theta\frac{\partial}{\partial \theta}+\frac{1}{r^2 \sin^2\theta}\frac{\partial^2}{\partial \phi^2},
\end{eqnarray*}
\noindent
$\gamma = \beta+\nu, \rho = \nu - \beta,$ and $\mu = \cos\theta.$ We refer \cite{kmt} for the explicit expressions of the source terms $S_{\rho}$ $S_{\gamma}$ and $S_{\omega}$. These basic differential equations can be transformed into an integral representation to handle the boundary conditions in relatively simple manner with the help of multi dimensional Green's functions and solved by Komatsu-Eriguchi-Hachisu method\cite{kmt,kmt1}.

\par
A maximum limit for the stable rotation of a star $\Omega_{K}$, is set by the onset of mass shedding from the equator of the star. General relativistic expression for this Keplerian frequency `$\Omega_{K}$' can be obtained by using the extremal principle to the 
circular orbit of a point particle rotating at the equator of the star \cite{NKGFB}.
This leads to the expression for the Kepler frequency, which is given as, 

\begin{equation}
\Omega_{K}= \omega + \frac{\omega^{'}}{2\psi^{'}} + e^{\nu-\beta}\left
[\frac{1}{R^2}\frac{\nu^{'}}{\psi^{'}} +\left(\frac{e^{\beta-\nu}\omega^{'}}
{2\psi^{'}} \right)^2  \right]^{1/2},
\end{equation}
\noindent

where, $\psi=\beta^{'}+\frac{1}{R}$, and $'$ denotes the differentiation with
respect to the radial co-ordinate $r$.

\par
Next, the general relativistic expression for the moment of inertia $I$ of a 
rotating star is given by \cite{hartless}

\begin{equation}
I=\frac{1}{\Omega}\int dr d\theta d\phi \sqrt{-g}T^{t}_{\phi},
\end{equation}

where $g$ is the determinant of the metric (\ref{metric}). Using eqn. (14-16), the
moment of Inertia is then given as,

\begin{equation}
I=\frac{2\pi}{\Omega} \int dr \int d\theta~ r^{3} \sin^{2}\theta \frac{(P+\epsilon)v}{1-v^{2}}e^{2(\alpha+\beta)},
\end{equation}

and hence the angular velocity `$J$' of the star can be obtained from the relation

\begin{equation}
J=I\Omega. 
\end{equation}

\par
We use the code written by Stergioulas \cite{steir} based on the Komatsu-Eriguchi-Hachisu method to construct uniformly rotating star models.

\section{Results and discussions}
\subsection {Choice of parameters and compatibility with the Heavy-Ion collision data}

From our earlier work \cite{tkj06} it was conclusive that
the effect of varying incompressibility on the underlying nuclear equation of state
and the star properties is minimal. There we found that EoS with incompressibilities 
$K=210, 300$ \& $380~ MeV$ resulted in stars with nearly same maximum mass 
at similar central densities. Similar results were obtained in other global properties of the compact star such as gravitational redshift,
maximum baryonic mass and radius too. However the difference in nucleon effective mass
was found to be prominent both in the dense matter EoS and also on the star properties. 
We now investigate the effect of varying nucleon effective mass on the 
rotational attributes of the evolved neutron star using the same model 
and same set of parameters. It is to be noted that the parameter set is in accordance with recently obtained heavy-ion collision data \cite{daniel02}. Therefore in the
present context it shall be interesting to see the implications of the constraints
from heavy-ion collision data on neutron star structure. 

Here it is imperative to mention that we chose the parameters for the present analysis
keeping the incompressibility ($K=300 MeV$) constant. The incompressibility is higher in comparison to other successful mean field models like the NL3 ($K=271.5 MeV$) \cite{nl3}. 
However, is well known that the nuclear symmetry energy greatly influences the EoS at high densities.
The value for symmetry energy $J =37.4 MeV$ in case of NL3 is unphysically large in
comparison to that we have in the present model ($J =32 MeV$). On the other hand the
DBHF, which is considered to be more realistic EoS in the non-relativistic domain 
has a symmetry energy of $J =31.5 MeV$ and is comparable to our prescription 
and is in good agreement with the empirical models and also with the variational 
approach \cite{dbhf}. In addition to these aforesaid differences in the nuclear saturation properties, there are variations in the nucleon effective mass in various models.
We take $m_n^{\star}/m_n~=~(0.80~-~ 0.90)$ which is consistent with the values
obtained from the analysis of neutron scattering off lead nuclei \cite{mr,nuclei}. 
  
\begin{figure}[ht]
\vskip 0.3in
\begin{center}
\includegraphics[width=7cm,height=7cm,angle=0]{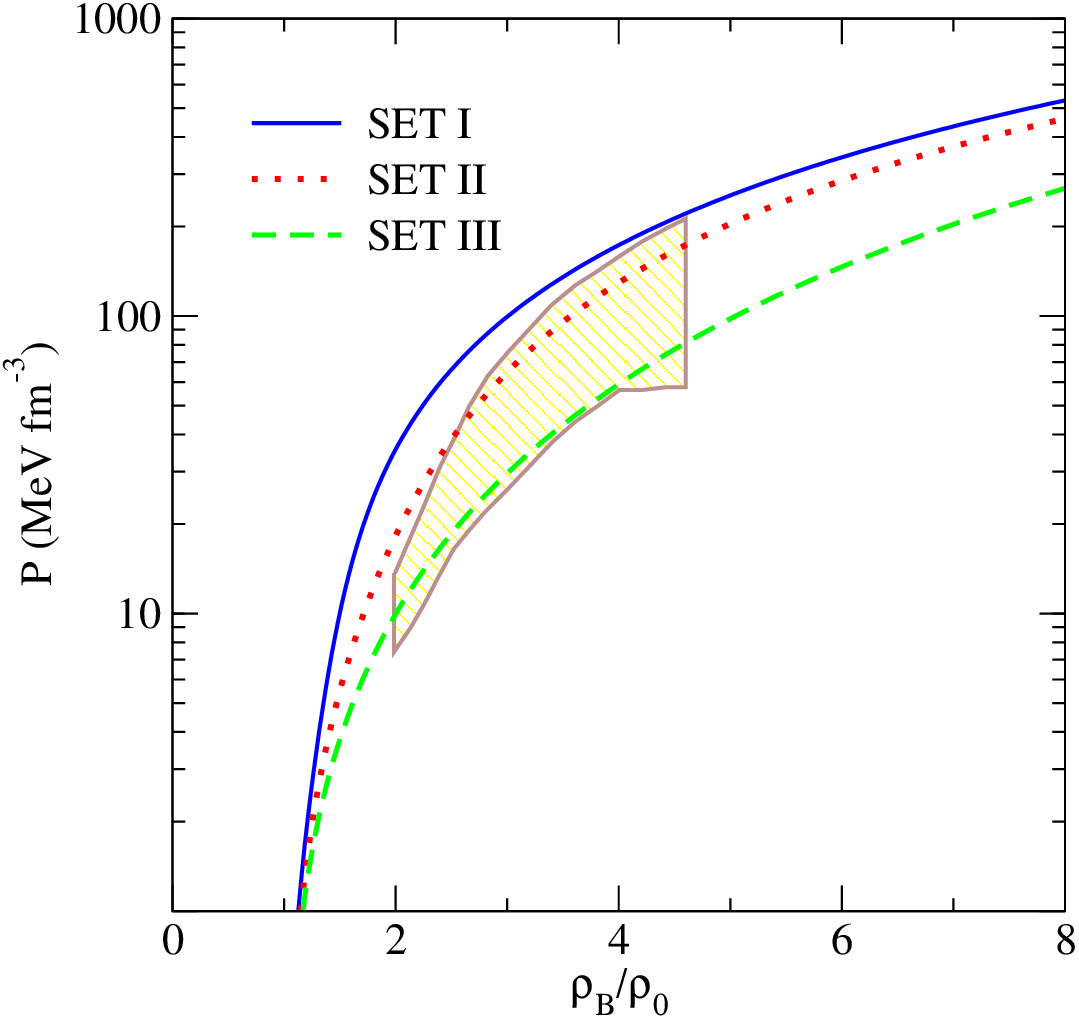}
\end{center}
\caption
{Pressure ($MeV fm^{-3}$) as a function of normalized baryon density ($\rho_B/\rho_0$) 
of symmetric nuclear matter. The shaded region is consistent with the experimental 
flow data \cite{daniel02}.}
\end{figure}

\begin{figure}[ht]
\begin{center}
\includegraphics[width=7cm,height=7cm,angle=0]{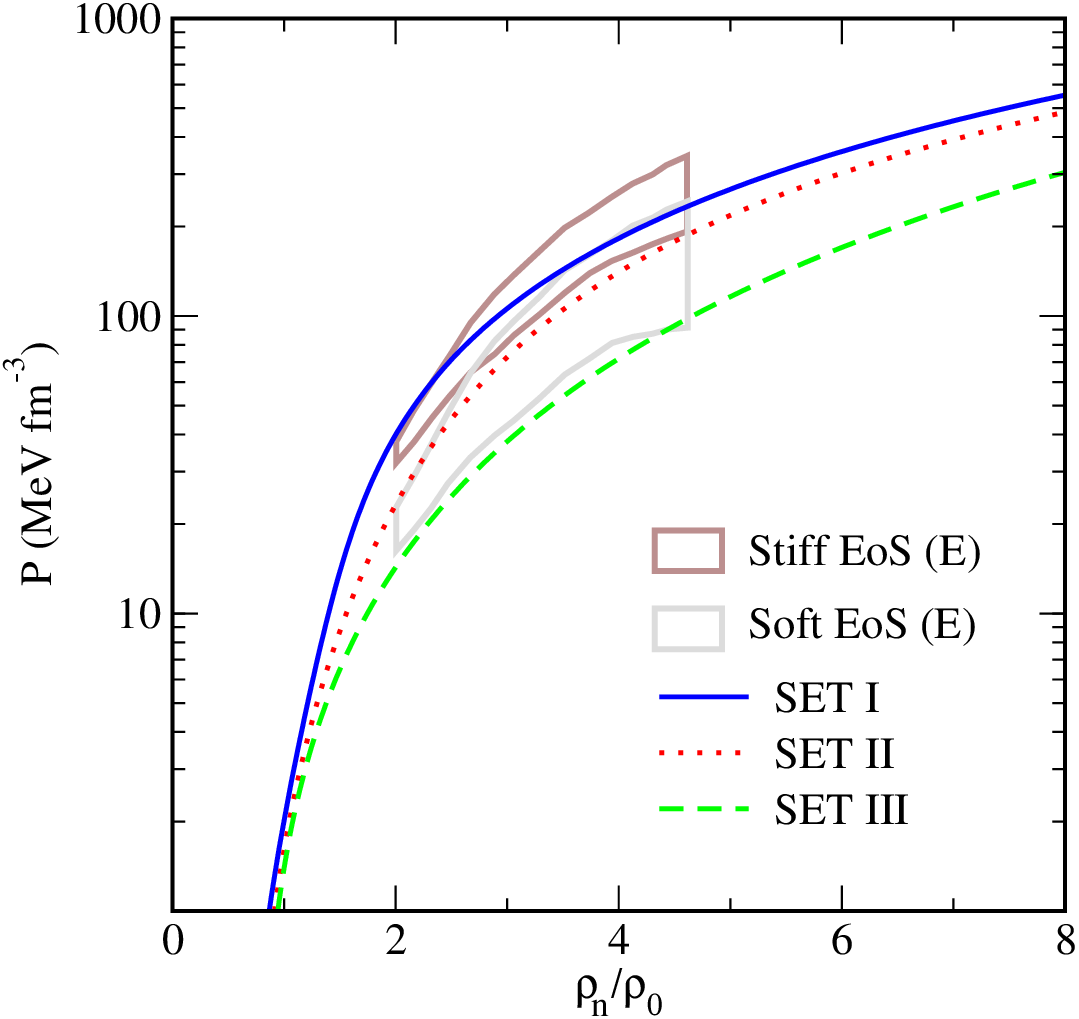}
\end{center}
\caption
{Pressure ($MeV fm^{-3}$) as a function of normalized baryon density ($\rho_n/\rho_0$) 
of neutron matter. The shaded region is consistent with the experimental 
flow data \cite{daniel02}. The upper polygon represents the stiff EoS, and the lower
one corresponds to data for a soft EoS.}
\end{figure}

The parameter set for the present work is listed in Table-1,
which vary in nucleon effective masses $(m_N^{\star}/m_N = 0.80, 0.85~\&~ 0.90)$,
but has the same incompressibility value at $K = 300 MeV$.
The parameter sets satisfies the nuclear saturation properties,
$E_B$, energy per nucleon, $-16.3 $ MeV at saturation density $0.153~ fm^{-3}$,
effective nucleon Landau mass $(0.8-0.9) ~m_N$, incompressibility $K=300 MeV$,
and asymmetry energy coefficient value ($\approx~ 32$ MeV), so that our
extrapolation to higher density remains meaningful. The coupling constant 
$c_{\rho N}$ is fixed to reproduce $a_{\rm sym}$
to the empirical value, 32 $\pm$ 6 MeV\cite{moll88}.
This gives $c_{\rho N}=4.66~ \hbox{fm}^2$ for $a_{\rm sym}$=32 MeV.
The first three columns in table 1 gives the coupling strength of the 
meson fields and the nucleon, namely the scalar field $c_{\sigma N}$, the vector 
field $c_{\omega N}$ and the iso-vector coupling strength $c_{\rho N}$, where
$c_{i N}~=~(g_{i N}/m_{i})^2$, for $i~=~\sigma, \omega, \rho$. 
The next two columns are the higher order 
scalar field constants 'B' and 'C', required to reproduce the bulk nuclear 
matter saturation properties with reasonable incompressibility which is 
given in the next column, followed by the nucleon effective mass for the three
sets. With these parameterization, $m_\sigma = m_N \sqrt{(c_{\omega N}/c_{\sigma N})}$, 
turns out to be in the range (509 - 627)MeV for the three parameter sets enlisted in Table I.
Note that in these calculations, other meson masses or the respective couplings
do not enter separately but through the ratio's $c_{\rho N} = (g_{\rho N}/m_{\rho})^2$
and $c_{\omega N} = (g_{\omega N}/m_{\omega})^2$.

It is to be noted that the nucleon effective mass taken in the present work
lies in the range $m^{\star}_N/m_N= (0.80-0.90)$. These values are considerable larger 
as compared to the Walecka model. A higher effective nucleon mass is known
to generate problems in describing finite nuclei properties. However, these
values are consistent with the results derived from non-relativistic
analysis of scattering of neutrons from lead nuclei \cite{nuclei}. 
We shall show that the variation in nucleon effective mass shall have considerable impact 
on the underlying EoS of the dense matter regarding its composition and hence on the
gross stellar properties. As stated earlier that the EoS considered for the present
study is in accordance with recently obtained heavy-ion collision data. In Fig. 1, we
compare our EoS with the experimental flow data from recent heavy-ion
collision for symmetric nuclear matter \cite{daniel02}. Here we find that 
among the three parameter set set II and III are in very good agreement within
the entire density domain of $\approx(2-5)\rho_0$. Similarly in Fig. 2, we compare the 
case for neutron matter with the experimental flow data. The two shaded region correspond to the
density dependence of symmetry energy. The upper one corresponds to a strong density
dependence (stiff EoS) while the lower one corresponds to a weak density dependence
(Soft EoS) of the symmetry energy \cite{daniel02,pra}. Here, all the three 
EoS seems to be in good agreement with the experimental data. However set I seems 
represents the stiffest EoS among the three and set III is the softest prescription. 
Therefore, in the present context, it can be seen that the flow data favors a soft EoS. 
 
\begin{table}
\caption{Parameter sets for the model.}
\vskip 0.1 in
\begin{center}
\begin{tabular}{cccccccccccc}
\hline
\hline
\multicolumn{1}{c}{set}&
\multicolumn{1}{c}{$c_{\sigma N}$} &
\multicolumn{1}{c}{$c_{\omega N}$} &
\multicolumn{1}{c}{$c_{\rho N}$}   &
\multicolumn{1}{c}{$B$} &
\multicolumn{1}{c}{$C$} &
\multicolumn{1}{c}{$K$} &
\multicolumn{1}{c}{$m_N^{\star}/m_N$} \\
\multicolumn{1}{c}{ } &
\multicolumn{1}{c}{($fm^2$)} &
\multicolumn{1}{c}{($fm^2$)} &
\multicolumn{1}{c}{($fm^2$)} &
\multicolumn{1}{c}{($fm^2$)} &
\multicolumn{1}{c}{($fm^4$)} &
\multicolumn{1}{c}{($MeV$)}  &
\multicolumn{1}{c}{}\\
\hline
I   &8.5   &2.71  & 4.66  &-9.26  &-40.73  &300  &0.80 \\
II  &6.79  &1.99  & 4.66  &-4.32  &0.165   &300  &0.85 \\
III &2.33  &1.04  & 4.66  &9.59   &46.99   &300  &0.90 \\
\hline
\end{tabular}
\end{center}
\end{table}

\subsection{Composition of Hyperon rich matter}

The composition of charge neutral dense matter is very sensitive to the
hyperon-meson coupling parameters which however are very poorly known \cite{prak97,nkg01}.
There are different phenomenological prescriptions e.g., using the quark counting
arguments based on SU(6) theory to fix the vector couplings, while, fixing
the scalar couplings from the potential depths of hyperon in nuclear matter
\cite {form1,mishra,su6}.  
From various analysis the choice of the ratio of hyperon to nucleon coupling
for $\sigma$ meson $x_\sigma < 0.72$ has 
been emphasized \cite{form} and also from studies
based on hypernuclear levels \cite{rufa90}, the choice ($x_\sigma < 0.9$)
is bounded from above. Following this convention,
in our present work we take $x_{\sigma}$=$g_{\sigma H}$/$g_{\sigma N}$=$0.7$,
$x_{\omega}$=$g_{\omega H}$/$g_{\omega N}$=$0.783$ and $x_{\omega}$
=$x_{\rho}$, to calculate the EOS for the neutron star matter. 
This leads to the binding of $\Lambda^{0}$ in nuclear matter:
$(B/A)_{\Lambda}$=$x_{\omega}g_{\omega}\omega_{0}+m^{*}_{\Lambda}
-m_{\Lambda}\approx-30$ MeV. We further assume that couplings to the $\Sigma$
and $\Xi$ resonances are equal to those of the $\Lambda$ hyperon \cite{mr,form}.
It is worth noticing that taking the $\rho-$meson coupling as $x_{\omega}$
=$x_{\rho}$ or $x_{\sigma}$ =$x_{\rho}$ does not alter the EoS substantially. 
Further that among all the three coupling schemes, this choice restricts the 
equation of state of neutron star matter following the constraint of $\Lambda^{0}$ 
binding in nuclear matter, ably supported by the hypernuclei experiments \cite{hyper}.
We would like to mention here however that the values of these potential are
somewhat different \cite{diff} from the analysis of $\Sigma^{-}$ atomic data
and the final state interaction of $\Xi$ hyperons in experiments $E224$ at $KEK$
\cite{kek} as well as $E885$ at $AGS$ \cite{ags}, which indicates a repulsive potential
for $\Sigma^{-}$ hyperon in dense matter. However, the exact values of the relativistic
potential remain inconclusive. We shall carry out our numerical analysis
with the values of the potentials as prescribed above.

As stated earlier, the equation of state of dense matter is very 
sensitive to the nucleon effective mass.
In Fig. 3, we show the respective particle densities of a charge neutral hyperon rich matter
for $m^{\star}/m=0.80$ (set I) upto ten times nuclear matter density. Similarly 
Fig. 4 and Fig. 5 displays the
particle population for set II ($m^{\star}/m=0.85$) and set III 
($m^{\star}/m=0.90$) respectively. From the plots, the sensitivity of the nucleon 
effective mass on the respective particle composition is clearly reflected.
It can be seen that with increase in the nucleon effective mass the appearance of various
hyperons is pushed further to higher densities. The appearance or 
concentration of the hyperon species at a particular baryon density 
is known to regulate the degree of softening effect on 
the equation of state, which is then manifested in the global 
properties of compact stars. For example, parameter set I and set II accommodates
all the octet of baryons upto ten times normal nuclear matter density although
the hyperons starts appearing at higher densities in the later case. However in case of
set III, the nucleon chemical potential is not high enough to have $\Sigma^+$ and
$\Xi^0$ even upto highest of density region. From the systematics of the nucleon mass
in hyperon rich matter, we recall that set II is endowed with repulsive force 
dominance unlike set III, which is dominated by the scalar forces \cite{tkj06}.

\begin{figure}[ht]
\vskip 0.3in
\begin{center}
\includegraphics[width=7cm,height=7cm,angle=0]{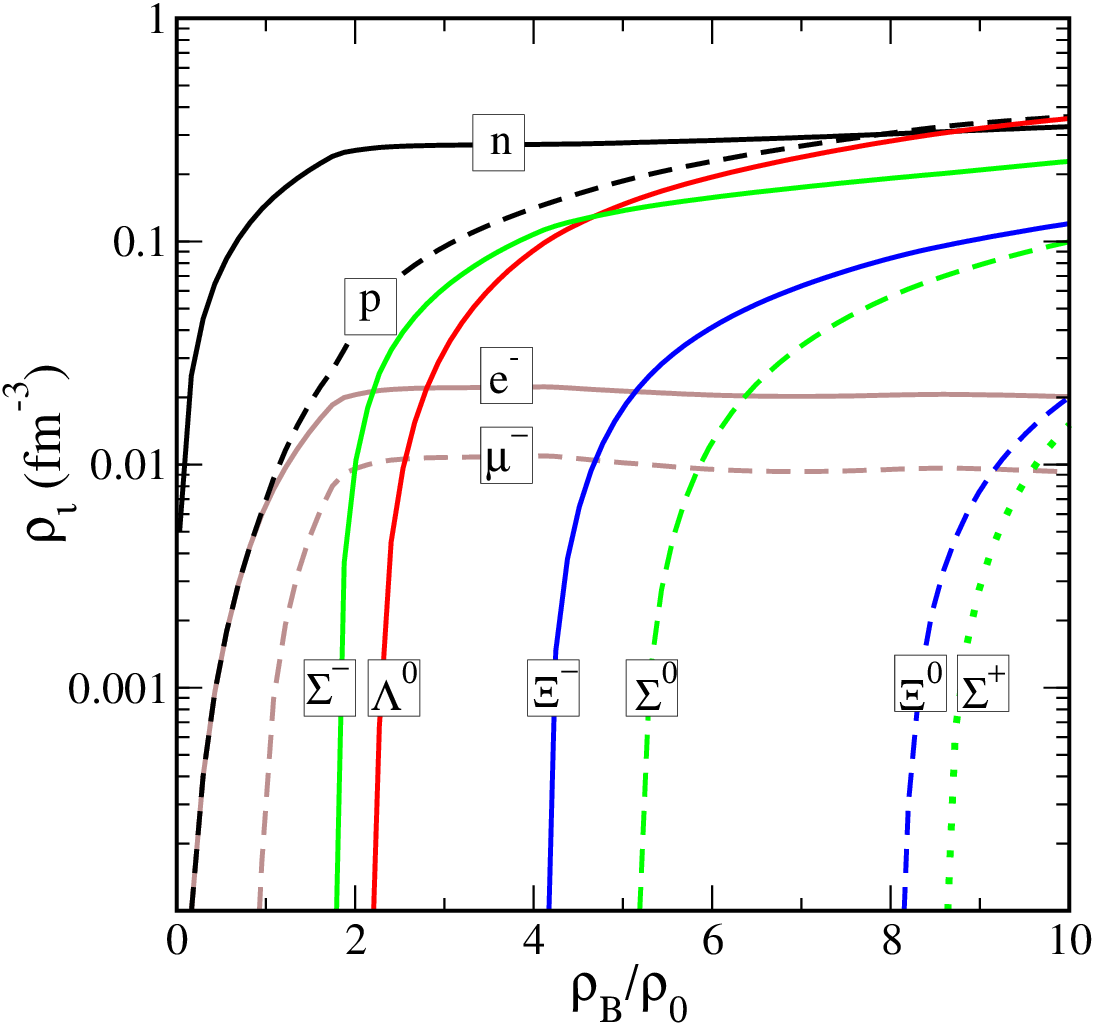}
\end{center}
\caption
{Particle densities ($fm^{-3}$) in neutron star matter for parameter set I.}
\end{figure}

\begin{figure}[ht]
\begin{center}
\includegraphics[width=7cm,height=7cm,angle=0]{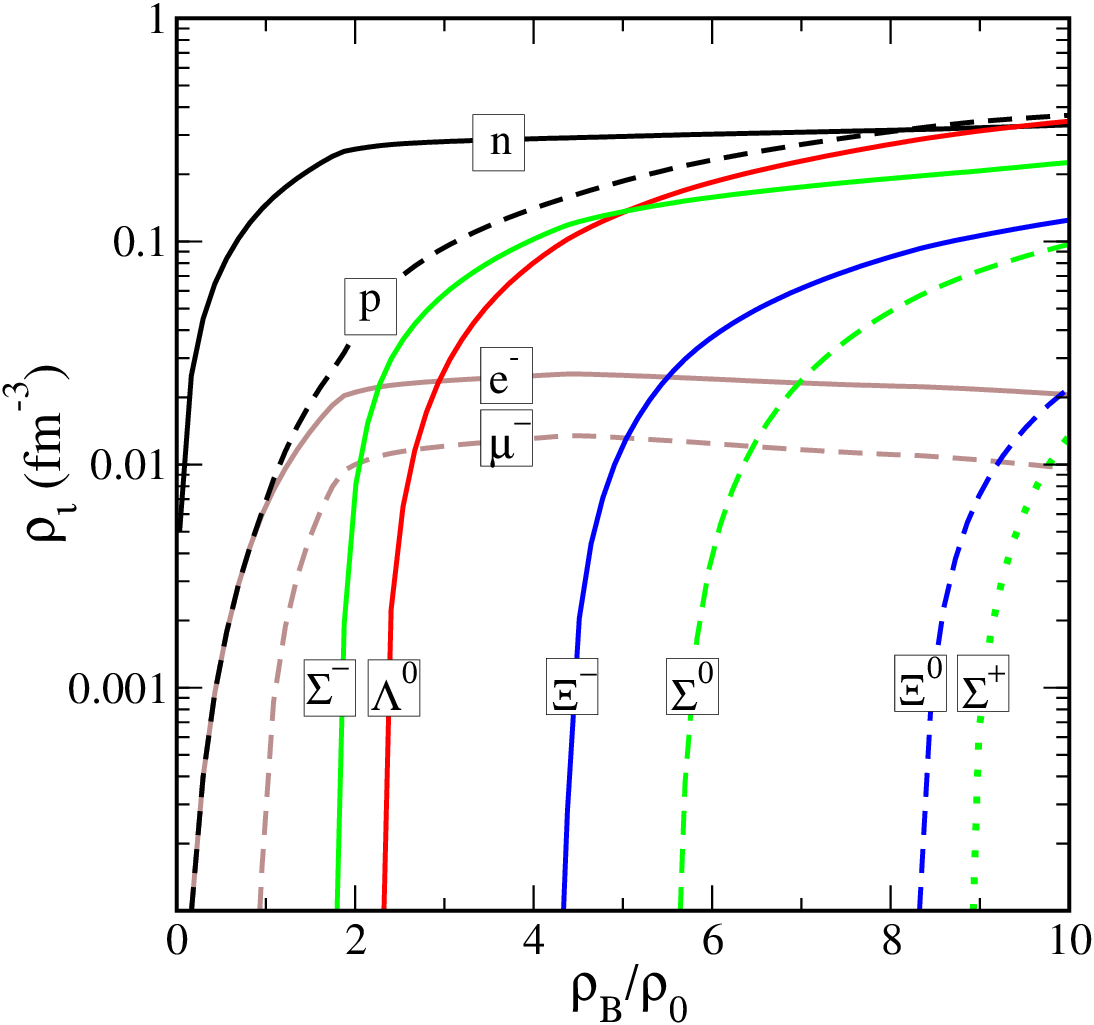}
\end{center}
\caption
{Same as Fig. 1 but for set II.}
\end{figure}

\begin{figure}[ht]
\vskip 0.3in
\begin{center}
\includegraphics[width=7cm,height=7cm,angle=0]{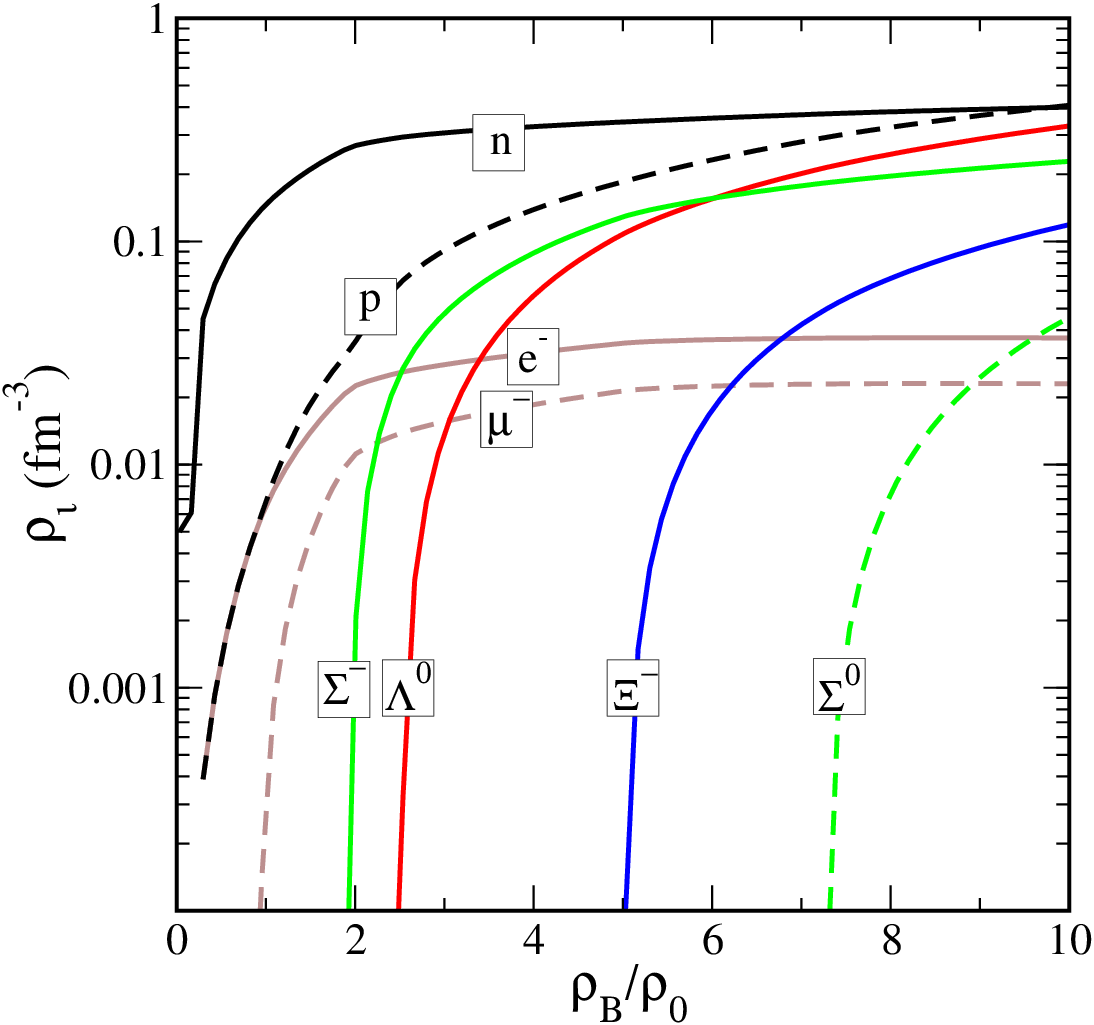}
\end{center}
\caption
{Same as Fig. 1 but for set III.}
\end{figure}

In general, for all the parameter sets, the hyperons starts appearing at 
$\approx 2\rho_0$ and are the dominant species in par with nucleons at 
$\approx 10\rho_0$. It can also be seen that the negatively charged particle
species are favored in the dense matter composition followed by the neutral ones and
the positively charged counterparts. For example in all the three cases, one finds that
$\Sigma^{-}$ appears before $\Lambda^{0}$ although the vacuum mass difference between the two is 
$\approx 77 MeV$ and the former is massive than the later. This is due to the fact that
as the density is increased ($\approx 2\rho_0$), the negatively 
charged $\Sigma^{-}$ starts competing
with the negatively charged leptons in maintaining charge neutrality of the dense matter. 
Due to the depletion of the lepton concentration the corresponding chemical potential 
$\mu_e$ decreases, thereby lowering the threshold chemical potential of $\Sigma^{-}$
as compared to $\Lambda^0$. Similar 
arguments exist for the other negatively charged baryons. 

\subsection{Global properties of a rotating Neutron star}

\begin{figure}[ht]
\vskip 0.3in
\begin{center}
\includegraphics[width=7cm,height=7cm,angle=0]{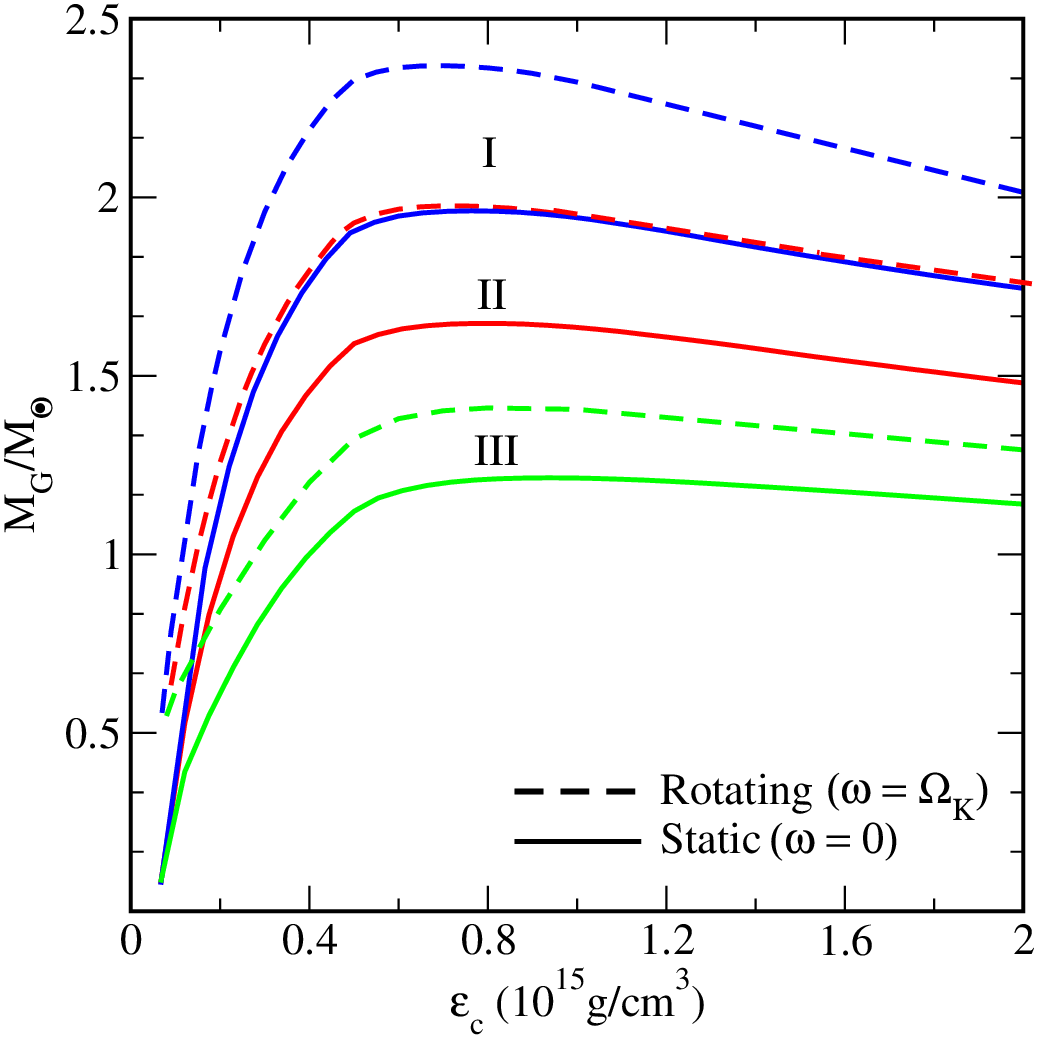}
\end{center}
\caption
{Dependence of Gravitational mass of the star $($in $M_{\odot})$ on the 
central density of the star $(10^{15}g cm^{-3})$ for the three cases in comparison
with the static results (solid lines).}
\end{figure}

The properties of the neutron star is unique to the EoS considered. Using these three EoS 
we now calculate some of the global properties of the neutron star, which is rotating
at its Kepler velocity $(\Omega_K)$ by employing the $RNS$ code \cite{steir}, as
described in section III.

In Fig. 6, we compare the variation of gravitational mass of the star for the three 
cases as a function of the central density of the star for both the static
and the rotating one. In later case, we considered rotation till the onset of 
mass shedding at the equator or equivalently upto its Kepler frequency.
The effect of rotation on the stellar mass is clearly visible from the plot, where 
it is noticeable that a rotating star acquires more mass than its static counterpart.
For all the three case at $(\omega = \Omega_K)$, we notice an increase of nearly
17-20 \% in the star mass in comparison to the static ones. However, this increase in mass doesn't alter the central density of the star very much. The maximum mass 
obtained for set I, II and III, for the rotating case are
$2.4 M_{\odot}$, $1.9 M_{\odot}$ and $1.4 M_{\odot}$ respectively and the
central densities obtained for both set I and II are $7.0 \times 10^{14}g cm^{-3}$ 
and  for set III, it is $8.0 \times 10^{14}g cm^{-3}$. Thus we find that for the three cases
the maximum mass is obtained at a central density of $(2.5 - 3)\rho_0$. In case of
non-rotating star the central density obtained at maximum mass lies in the range 
$(3 - 3.5)\rho_0$. The corresponding maximum baryonic masses for set I, II and III for the rotating case are 2.56$M_{\odot}$, 2.11$M_{\odot}$ and 1.47$M_{\odot}$ respectively.  
The difference between the gravitational mass and the baryonic mass gives us the
gravitational binding of the star. The fact that at any given density the baryonic mass
exceeds the gravitational mass of the star is a typical of compact stars.
Consequently, we find that the EoS with lower nucleon effective mass results in
more bound compact systems. From the observational point of view, we find that
recent observations of neutron star masses like $M_{J0751\pm1807}$
=2.1$\pm$0.2 $M_{\odot}$\cite{mass1}, $M_{4U 1636\pm536}$=2.0$\pm$0.1 $M_{\odot}$\cite{ns02},
$M_{Vela X-1}$=1.86$\pm$0.16 $M_{\odot}$\cite{ns-ob7} and
$M_{Vela X-2}$=1.78$\pm$0.23 $M_{\odot}$\cite{ns04,ns-ob8} predicts massive stars.
Our results are in very good agreement with these observations, except for set III which
results in low mass star among the three prescription and has the highest 
nucleon effective mass value. It is worth to recall that although there are 
large observational errors in the mass-radius determination, yet in case of 
$Vela X-1$ the lower mass limit $\approx (1.6-1.7)M_{\odot}$ is atleast mildly
constrained by geometry \cite{geo}.
  
\begin{figure}[ht]
\vskip 0.3in
\begin{center}
\includegraphics[width=7cm,height=7cm,angle=0]{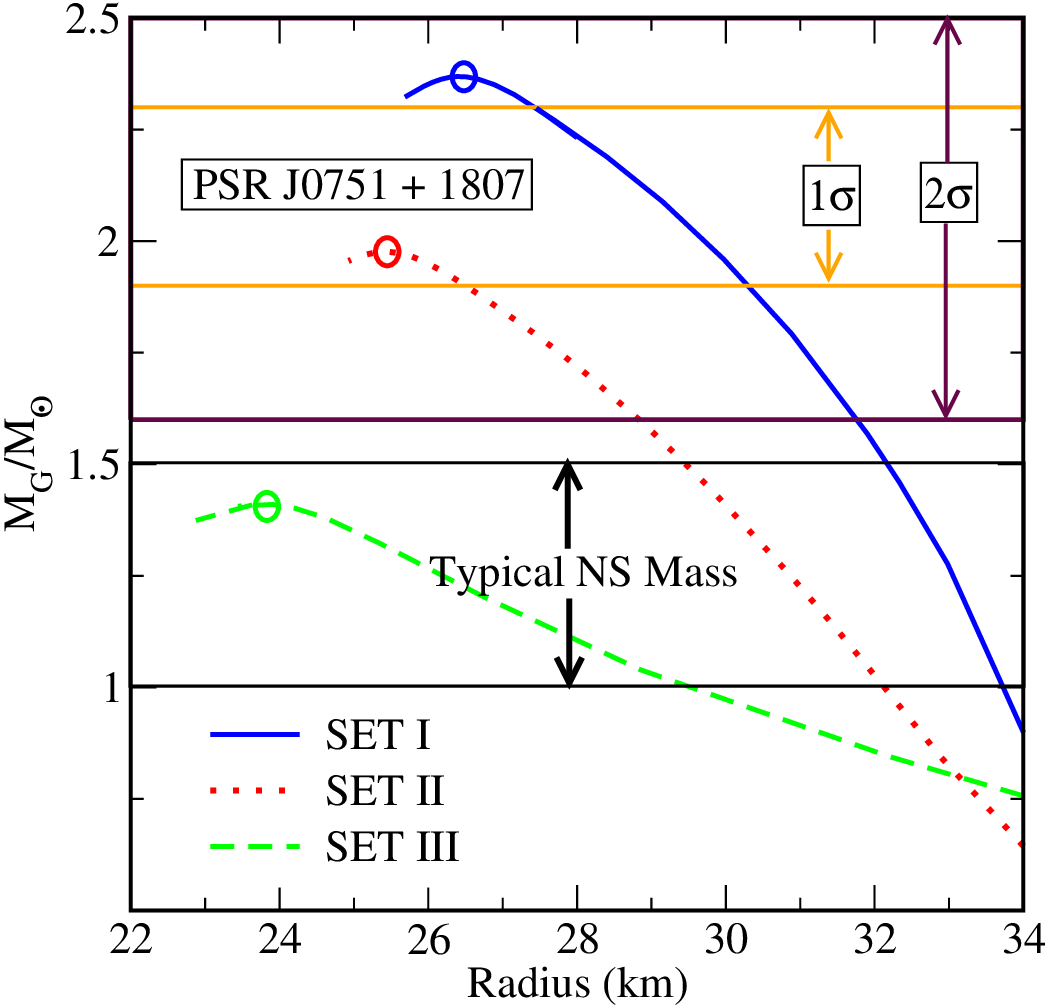}
\end{center}
\caption
{Gravitational mass of the star ($M_{\odot}$) rotating at Kepler frequency as a function of  
the circumferential radius of the star (in $km$). Also plotted is the 
typical neutron star mass range $(1.0 - 1.5)$ $M_{\odot}$ and the mass estimates 
obtained from $PSR J0751 + 1807$ with error estimate of $1\sigma$ 
($2.1 \pm 0.2$)$M_{\odot}$ and $2\sigma$ (${2.1}_{-0.5}^{+0.4}$)$M_{\odot}$ \cite{mass1}.}
\end{figure}
 
\begin{figure}[ht]
\begin{center}
\includegraphics[width=7cm,height=7cm,angle=0]{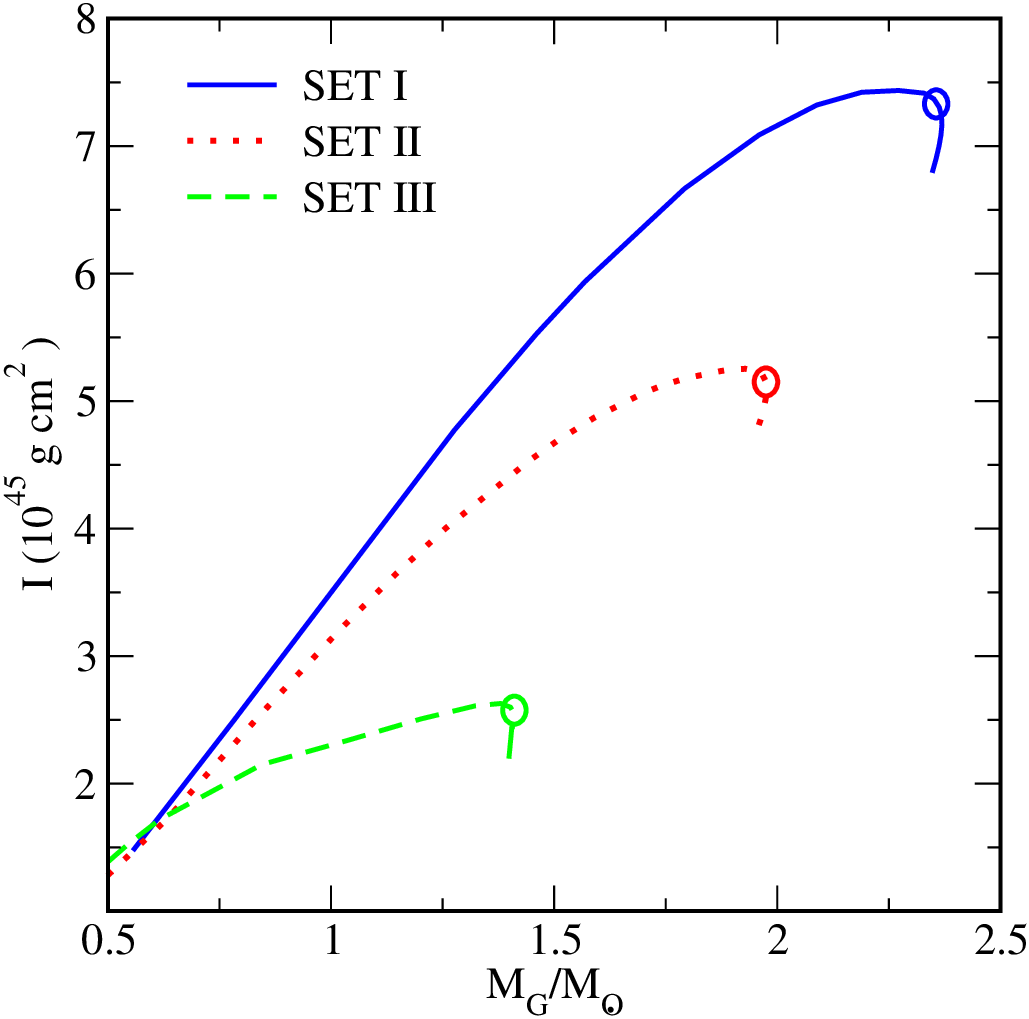}
\end{center}
\caption {Moment of Inertia ($10^{45} g cm^2$) of the star as a function of  
the mass of the star. The circled point denotes the value at maximum mass configuration.}
\end{figure}

Of the three EoS, in the present analysis, parameter 
set I is the stiffest prescription and set III is the softest one.  
This property is well documented in Fig. 7, where we plot the mass of the neutron
star as a function of the radius and compare our results with the typical observed mass
of neutron stars ($1.0 - 1.5$)$M_{\odot}$ and also with more massive ones such as
the case of $PSR~J0751 + 1807$ within the error estimates of $1\sigma$ and $2\sigma$.
Here we find that our results are in good agreement with these observations, precisely
set II and III agree with the estimates from $PSR~J0751 + 1807$. 
In order to evaluate the radius of the star, we have 
included the BPS EoS \cite{bps} at subnuclear densities which contributes to
the crustal part of the star structure. The radius obtained for the three EoS 
lies in the range ($24 - 26$) km and the 
compactness parameter ($M/R$) thus lies in the range ($0.05 - 0.09$)$M_{\odot}/km$.
From the observational point of view, there are large uncertainties in determination of
the radius of the star \cite{rad1,rad2,rad3} which is primarily because of our lack
of knowledge of the composition of the star atmosphere, large distance and also
due to the presence of high magnetic fields. On the other hand,
the general relativistic limit for the compactness parameter assuming a uniform
density star with the causal equation of state i.e., $P=\varepsilon$ gives
$M/R < 4/9$ \cite{mr}. In comparison, our prediction lies far below this limit
which reflects the softness of our EoS ably supported from the heavy-ion flow data (Fig. 2).
Consequently, the flattening parameter, which is defined as 
the ratio of the polar to the equatorial radius 
($R_p/R_e$), for all the three case is $\approx 0.59$, at $\Omega_K$.
Rotation induces deformation in the shape of the star
which leads to a dependence of the star's metric on the polar coordinate $\theta$.
In general, rotation stabilizes the star against gravitational collapse
and therefore rotating neutron stars are more massive than the static ones.
Further, the additional centrifugal forces in a rotating star help to 
counteract the pull of gravity, resulting in larger radii for a given mass.

Moment of inertia of neutron stars plays crucial role in the models of radio
pulsar. Independent of the rotation, i.e., slow or fast, the relation of moment of
inertia to the matter distribution within the star is complicated. Among all the global
properties of the star, the moment of inertia is the most sensitive one to the dense matter
equation of state. In Fig. 8, we show the variation of the moment of inertia 
of the rotating neutron star as a function of star mass for the three parameter sets. 
Here it can be seen that set I ($m^{\star}/m = 0.80$)
results in the maximum value of the moment of inertia at ($I_{45}~=~7.22$; $I_{45}~=~
10^{45} g cm^{2}$) and as we move towards the higher effective mass values, 
this decrease becomes increasingly prominent. But for all the cases we observe 
that moment of inertia of the compact stars increases rapidly as it approaches the
Kepler velocity or till the mass shedding limit. The moment of inertia obtained for set
II and III are $I_{45}~=~5.14$ and $I_{45}~=~2.54$ respectively.
From the observational point of view, recently discovered relativistic pulsar 
$PSR J0737-3039$ \cite{mi-pulsar} could be the first one in which the moment of 
inertia is measured. Also the estimates from crab pulsar \cite{crab} puts
$I_{45} > 1.61$ for $M_{neb}=2.0 M_{\odot}$, which is the conservative estimate,
and $I_{45} > 3.04$ for $M_{neb}=4.6 M_{\odot}$ (latest estimate). 

\begin{figure}[ht]
\vskip 0.32in
\begin{center}
\includegraphics[width=7cm,height=7cm,angle=0]{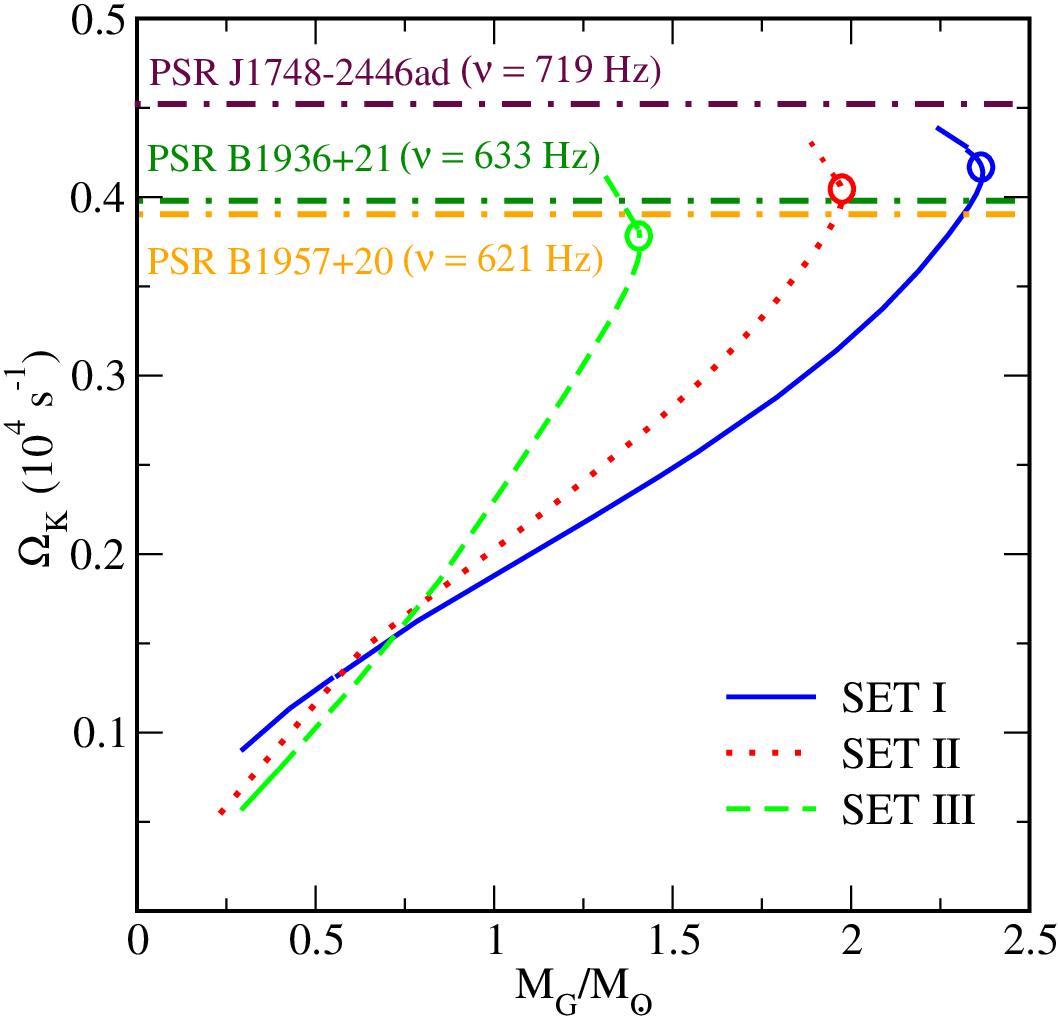}
\end{center}
\caption {Keplerian velocity $\Omega_K$ ($10^{4} S^{-1}$) as a function of  
the Gravitational mass of the star. The dot-dashed horizontal lines denotes the
observed frequency from the pulsars.}
\end{figure}

\begin{figure}[ht]
\begin{center}
\includegraphics[width=7cm,height=7cm,angle=0]{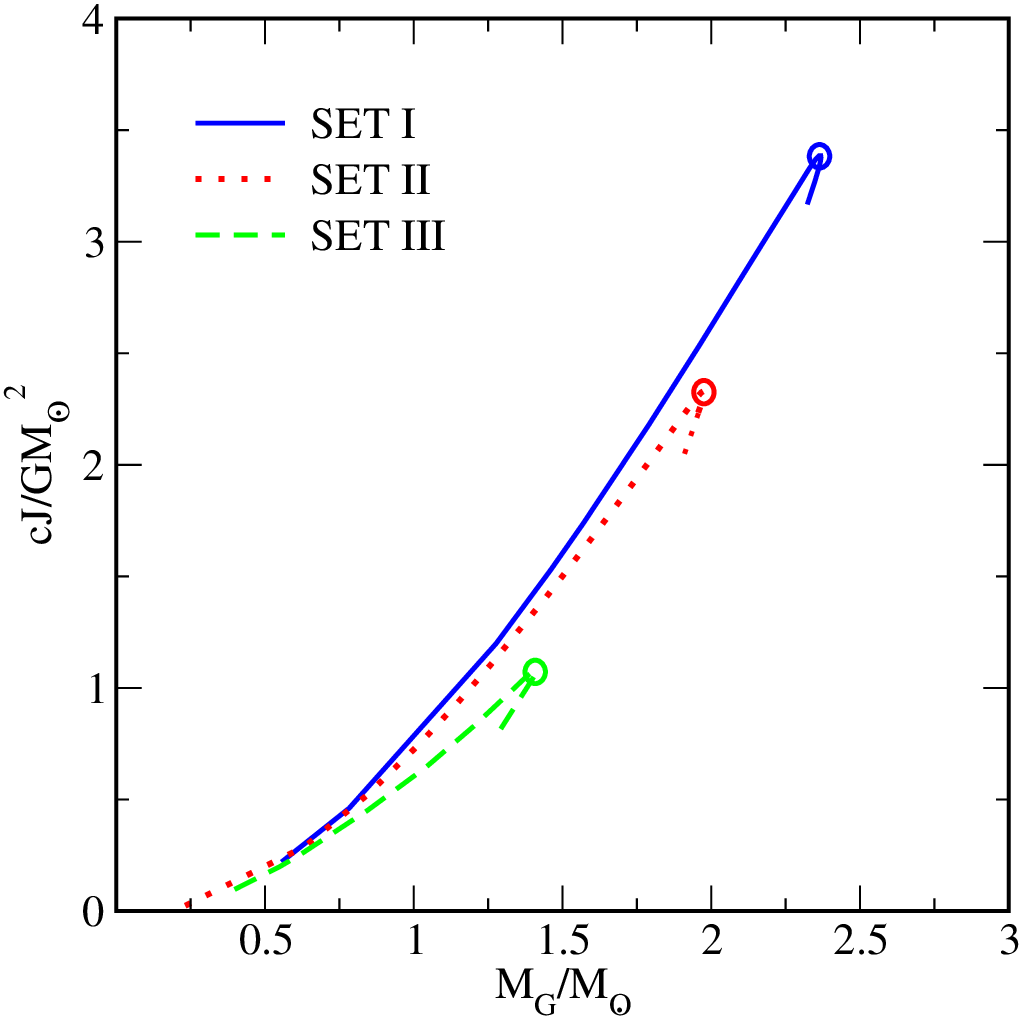}
\end{center}
\caption
{Angular momentum of the star as a function of the star mass.}
\end{figure}

In Fig. 9, we plot the Kepler angular velocity ($\Omega_K$) of the star as a 
function of the mass of the star. The Kepler angular velocity obtained for the three 
cases are $4141~ s^{-1}$,  $3999~ s^{-1}$ and $3714~ s^{-1}$ for set I, II and III 
respectively, which corresponds to a frequency of $659$Hz, $636$Hz and $591Hz$. 
The Kepler period, which is defined as $P_K=2\pi/\Omega_K$, obtained are
$1.52ms$, $1.57ms$ and $1.69ms$ for set I, II and III respectively. It is to be noted
that the values that we obtained presently with the three EoS is in agreement 
with the maximum observed value of the Kepler period for the observed fast rotating pulsars, 
namely the $PSR B 1937 + 21$ ($\nu~=633~Hz$) \cite{633} and $PSR B 1957 + 20$ 
($\nu~=621~Hz$) \cite{621}. On the other side, the constraint imposed by 
general relativity on the period of a 
relativistic compact star emphasize $P > 0.24~ ms$ \cite{mr}.
However there are indications of even faster rotating pulsars from the observational point of
view. Recently the observation of the fastest rotating neutron star, with a frequency
$\nu~=~1122~ Hz$ in an X-ray burst from the X-ray transient, $XTE~J1739-285$ was reported
\cite{rapid} in addition to the observation of $\nu~=~716~ Hz$ from $J1748-2446ad$ 
\cite{716}, which may set interesting insight into the structural aspects of
neutron stars. It has been mentioned that for the rapid rotation regime, one needs 
submillisecond pulsars with $\nu~>~1000~Hz$ \cite{cook}. However, the smaller
$\Omega_K$, that we get has to do with the large radii that we obtained in the 
present EoS. Therefore the present EoS while consistent with the observed 
$PSR B 1937 + 21$ and $PSR B 1957 + 20$ data, could not account for the pulsar rotating with
high frequency such as $\nu~=~716, 1122~ Hz$. This situation seems to be understandable
for the fact that the EoS employed presently represents the class of softest EoS in
comparison to other theoretical models that can explain the existence of fast rotors.
In this context, EoS for skyrmionic models appear to be promising \cite{jaikumar}.
However, with a phase transition structure possibly having a crystalline 
color superconducting quark matter may account for the fast 
rotating pulsars \cite{arxive}, before the mass-shedding limit. 

Similar results were obtained for the angular momentum of the star for the three cases, which
is shown in Fig. 10. It can be seen that the angular momentum increases rapidly
as the star reaches the mass-shed limit and finally attains a maximum value at maximum mass,
which is shown with the filled circles for the three cases.  

Finally, for a better correlation with the underlying nucleon effective mass, 
the overall global properties for the static and the rotational case for the three 
EoS is tabulated in Table 2.
\begin{table}
\caption{Comparison of the static (upper row) and the corresponding rotational 
(middle row) attributes of the neutron star predicted with the three parameter 
sets of the present model.}
\begin{center}
\begin{tabular}{cccccccccccccc}
\hline
\hline
\multicolumn{1}{c}{EoS} &
\multicolumn{1}{c}{$m^{\star}/m$} &
\multicolumn{1}{c}{$M$} &
\multicolumn{1}{c}{$R_{eq}$} &
\multicolumn{1}{c}{$\varepsilon_{c}/\varepsilon_{0}$} &
\multicolumn{1}{c}{$M_{b}$} &
\multicolumn{1}{c}{$I_{45}$}&
\multicolumn{1}{c}{$P_{K}$}\\
\hline
\multicolumn{1}{c}{} &
\multicolumn{1}{c}{} &
\multicolumn{1}{c}{($M_{\odot}$}) &
\multicolumn{1}{c}{}($Km$) &
\multicolumn{1}{c}{} &
\multicolumn{1}{c}{($M_{\odot}$)} &
\multicolumn{1}{c}{($10^{45} g cm^2$)}&
\multicolumn{1}{c}{($ms$)}\\
\hline
\hline
I      &0.80  &1.96   &17.5   &2.85  &2.18 &--   &--   \\
       &      &2.36   &26.4   &2.60  &2.56 &7.22 &1.52 \\
\hline
II     &0.85  &1.65   &16.7   &2.96  &1.81 &--   &--   \\
       &      &1.98  &25.5   &2.60  &2.11 &5.14 &1.57 \\
\hline
III    &0.90  &1.21  &15.0   &3.45  &1.31 &--   &--   \\
       &      &1.41  &23.7   &2.96  &1.47 &2.54 &1.69 \\
\hline
\end{tabular}
\end{center}
\end{table}

\section{Summary and outlook}

In our present work, we intended to present a unified approach to correlate the properties 
of dense matter EoS with respect to the variation in nucleon effective mass ($m^{\star}/m~=
~0.80, 0.85, 0.90$) defined at saturation density and recently obtained heavy-ion collision 
data \cite{daniel02}. Subsequently we study the rotational attributes of the sequence 
of neutron stars in a relativistic framework 
in the mean-field approach using an effective chiral hadronic model generalized to include
all the octet of baryons. The EoS employed in our present investigation satisfies
the nuclear matter saturation properties at reasonable incompressibility ($K=300~MeV$). 
We then compared the resulting EoS for the three parameter 
sets with the experimental flow data for both symmetric nuclear matter and neutron matter. We found that all the three parameter sets are in good
agreement with the flow data, but on a critical look, parameter set II $(m^{\star}/m~=~0.85)$ satisfy these two combined constraint quite well, however all the parameter sets
rather lie on the softer domain of neutron matter collision data. The resulting 
hyperon rich matter was also found to be very sensitive to the difference in the 
nucleon effective mass. From our analysis, it was conclusive that hyperons form 
a sizeable population in the dense matter and with higher nucleon effective mass value 
the threshold for appearance of the respective hyperon species is pushed further to
the higher densities. Hyperons are found to have a substantial impact on the density region
relevant to neutron star properties and is known to decrease the maximum mass of the star. The
ongoing and future experimental hypernuclear programs such as those at Jefferson Lab,
KEK, J-PARC and GSI, Darmstadt and FAIR etc. will provide the decisive insights into
the role of these exotic forms of matter in dense matter EoS and hence neutron stars, so
as to efficiently constrain the dense matter EoS.

We calculated the global properties of the rotating neutron star sequences such as the mass, radius, central density, moment of inertia and Kepler angular velocity in the KEH method or
the fast rotation approach. The maximum mass obtained at Keplerian velocity for the three sets
lies in the range $(1.4-2.4) M_{\odot}$ which is in very good agreement with recent
observations of massive neutron stars such as $PSR~J0751 + 1807$, although we
do find substantial increase in the star radius. On overall analysis, parameter set
II seems to fit the observed mass of $PSR~J0751 + 1807$ ($2.1\pm 0.2$) within 
1$\sigma$ error bar, which can be considered as ideal parameterization in the
present context. The compactness parameter ($M/R$) 
that we obtained for the three EoS represents the
lower limit from the observational point of view. The corresponding central
density at maximum mass lies in the range $(2.6 -3.0) \rho_0$. The flattening 
parameter i.e., the ratio of the polar radius to the equatorial radius obtained
for the three sets is $\approx~ 0.59$ which reflects the degree of deformation
that can result in case of rapidly rotating stars. Subsequently the circumferential 
radius obtained for the three cases lies in the range $(23.7-26.4)Km$. Further we
obtained the Kepler period in the range $(1.52-1.69)ms$, which can account for the 
fast pulsars observed such as the case of $PSRB1937+21$ and $PSRB1957+20$, rotating with
frequency of $\nu~=~633 Hz$ and $\nu~=~621 Hz$ respectively. However our EoS could not
explain the pulsars rotating beyond $\nu~=~659 Hz$, keeping in mind that we investigated
the neutron star sequences with a hyperon core. Possibly in order to account for pulsars
rotating at still higher frequencies, we require a stiff EoS with a phase transition structure 
in the neutron star core as emphasized in Ref \cite{arxive}.

The observation of high mass stars has raised considerable amount of debate over 
the nature of EoS i.e., the stiffness or softness of the EoS at high densities.  
In the present context we would like to stress that the EoS obtained in the present 
model belongs to the softest prescription
in comparison to other theoretical models, as also evident from the analysis with
respect to the experimental flow constraint. We also want to emphasize that the present EoS
is compatible with the DBHF at low density and the phenomenological RMF-NL3 parameterization
at high densities \cite{tkj04}. Besides, the
analysis in the present work throws interesting insights with regard to the implications
of heavy-ion collision data on the neutron star structure. Here we see that parameter
set I and III represents the upper and the lower bound of the flow data in the symmetric 
nuclear matter case, although all the three sets satisfies the soft prescription of the
neutron matter domain. Apparently this was also evident from the global properties of the star
that we obtained in our present analysis. The present EoS results in small Kepler
frequency before the mass-shedding limit which is for the fact that we obtain a large circumferential radii for the neutron stars in the present model, which greatly influences the
rotational attributes of the star. However we would like to mention here that 
more systematic calculations are required before we arrive at some precise 
conclusions on the constraints of dense matter EoS and neutron star properties provided
by the present model. We explored here the model of neutron star with a hyperon core, 
it shall be interesting to study and analyze these results including the phase 
transition aspects such as the transition to superconducting quark matter in the 
neutron star core, which we shall take up in our next endevour \cite{tkj07}.

\end{document}